\documentclass[pt12]{article}
\begin{document}

\begin{center}

          Quantum and Classical Phase Space: Separability and Entanglement\\

                           Ajay Patwardhan\\ 

              Physics Department , St Xavier's College, Mumbai 400001,India \\

                        November 8th 2002\\

 \end{center}                       

 The formalism of classical and quantum mechanics on
 phase space leads to symplectic and Heisenberg group
 representations, respectively. The Wigner functions
 give a representation of the quantum system
 using classical variables. The correspondence between
 the classical and the quantum criterion of separability
 for the system is obtained in terms of these functions.
 Entanglement is generic and separability
 is special. Some applications are discussed in commonly
 occuring examples and possibly in exotic systems.

\section {}

Introduction

{The separability conditions in classical dynamics and quantum dynamics can be obtained with phase space methods. There is a correspondence between them. Separability occurs as a special case. Wigner distributions on phase space play an important role in giving the representation of the Heisenberg group. Entanglement effects are a growing subject of research;(References 1 to 17).

(A)  The classical phase space

 The classical dynamics gives rise to n torii in the integrable  part of phase space.When separability conditions are satisfied ,an action angle representation for cyclic,periodic variables can be constructed. The Hamilton Jacobi method is used to construct explicit solutions.

Analytic solutions are known for a restricted set of problems. Otherwise integrals in involution with the invariant symplectic structure can be written.  $\{{I_m}, {I_n}\}=\delta_{m,n} $ for the skew symmetric two form $ I_mI_n{dq_m}\wedge{dp_n} $.

 The Hessian matrix obtained from the second partial derivatives of the Hamiltonian is used  for fixed point stability analysis and numerical solutions are found, which can be periodic and chaotic in phase space. For 2 x 2  or more dimensions in phase space chaos is common.

The classical phase space is generally a union of disjoint subspaces characterised by a invariant, the Kolmogorov entropy, which equals zero on the integrable and is positive on the chaotic subspaces. For each set function on these subspaces a( ergodic) invariant measure for integration on phase space exists.These measures restricted to constant energy surfaces could be used to write the micro canonical ensemble, and by integrating over all these energy surfaces a canonical ensemble can be defined.

This measure could be defined as: [exp(Kolmogorov Entropy)]x(usual volume element), within the distinct subspaces of phase space . Since the entropies are additive within each component subspace, and independent for subspaces with null intersection; this choice for separability could be generally valid.

 While a significant number of integrable Hamiltonian dynamics systems exist, the generic situation is that both chaotic and integrable phase spaces occur.Hence obtaining independent dynamics for the n degrees of freedom of a classical system is something special. For two or more $ (q_i,p_i)$ this effect is seen in examples like Henon Heiles Hamiltonians.

While the symplectic group Sp(2,R) has group isomorphism with GL(2,R), the $Sp(n\geq3,R)$ is not uniquely represented as $ GL(n\geq3,R)$. Braid groups $ B_n$ arise, corresponding to intertwining trajectories generated by the dynamics in phase space.For $ n\geq2$ chaos is possible.

(B) The quantum phase space

In quantum systems the symplectic structure goes over to the Heisenberg group representations.What are the corresponding properties , such as chaos and separability? As shown in ( Ref.1 R.Simon); using the Peres Horodecki criterion for a $ n=2 $ system, there is a separability condition for the Wigner functions.

In dimension 2x2 and 2x3 $ \rho $ is Separable $ <=> $ Positive Partial Transpose, but in higher dimensions $ \rho$  separable $ =>$ Positive Partial Transpose. For tripartite or correlated three particle systems, biseparable cases, with direct product of a state of one particle, with a linear combination of the diagonal density matrix elements for the other two particles can be found. Genuine entanglement in terms of inequivalent classes of states exist.

 The total Hilbert space has states which cannot be expressed as direct product states of one particle states. For n=3 the GHZ state $\frac{1}{\sqrt{2}}( |000> +|111>) $ and the W state $ \frac{1}{\sqrt{3}}( |100> +|010> +|001>) $ are the only two equivalence classes. Generally  a Hermitian operator W is to be found such that $ Tr(W\rho ) \geq 0 $ for separability, and $ Tr (W \rho) < 0 $ iff $ \rho $ is entangled. 

 For $n\geq 3 $ this criterion can be generalised for gaussian distributions as a factorisability condition. Entanglement appears as a generic property. The separability condition is neccessary but not sufficient for the $ n\geq3$ (tripartite or higher) case.The seperability condition fails to be a neccessary one when Wigner functions other than Gaussian are used, even for dimension two.
}
\section{}
{Quantum Phase Space of a 2n Mode System :Construction of the Wigner Distribution Conditions

In a 2n mode system conveniently constructed and modeled on  2n  oscillators, the classical 2n torii represent a set of 2n oscillators in phase space.The bipartite nature can be retained by pairing off the variables ,which is a particular canonical transform, and is a subgroup of the symplectic group Sp(2n,R). In this special case the construction of the two mode system can be generalised to the 2n modes, giving similar results.

Let $ \zeta = (q_i,$ alternating   with $  p_i)$ be a 4n vector in phase space.

 Then  $ [\zeta_a ,\eta_b] =i\Omega_{ab} $ is the symplectic form, which equals  $ \zeta_1 \eta_2 -\zeta_2 \eta_1$ for a two mode system .  Generally for a  2n mode system it is given by

\begin{displaymath}
  \Omega = \left(\begin{array}{c|c}
J & O \\
\hline 
 O & J
\end {array}\right) ,\qquad   J= \left(\begin{array}{c|c}
 O& I \\
\hline
 -I & O
 \end {array}\right)  \end{displaymath}

 and I is the nxn identity matrix. For taking the partial transpose consider $ q_{i= 1....n} $ and $ q_{n+1,.....,2n} $ as they are, $p_ {i=1,2.....,n}$ as they are but  $p_{n+1,.....,2n} $ with their signs inverted.

 This is to implement the mirror inversion only on the half of the 2n variables of the p type. For example the pairing is $ p_1-- > +p_1$  and $ p_{n+1}-- > -p_{n+1}$.

The Wigner distribution on 2n + 2n variables is

$ W(q,p)  = \pi^{-2n}\int d^{2n} q' <q-q' |\rho | q+q' > exp(2i q' p)$ 
}
\section{}

{ Harmonic Analysis in Phase Space : Conditions for Entanglement

For the transformation $(p,q,t)(p',q',t') = (p + p',q + q',t + t'+ \frac{1}{2}(pq' -qp'))$ the density is defined as a convolution in phase space.

$ V(f,g)(q,p) = <\rho(q,p) f,g> =\int dy exp(2\pi i q y ) f(y +\frac{1}{2}p)g(y - \frac{1}{2}p)$ and the Wigner function is the fourier transform given by

$W(f,g)(\zeta ,\eta) = \int\int( exp(-2\pi i (\zeta p + \eta q)))( V(f,g)(p,q))dpdq$

Consider the transformation in the Heisenberg group $(\zeta ,s)(\eta,t) --> (\zeta +\eta,s+t,\frac{1}{2}\Omega(\zeta,\eta))$ with

 $ W(\zeta)W(\eta) = W(\zeta+\eta)exp(\frac{1}{2} i \Omega(\zeta,\eta))$ as the composition for the W functions

 Canonical transforms $ \Omega(A\zeta,A\eta) = \Omega(\zeta,\eta)$ give the invariance on phase space. $\zeta$ and $  \eta $ are '4n' dimensional vectors in phase space.

 $ \Omega(\zeta, \eta) = \eta^T J \zeta $ where ``J'' is the Jacobian ;$ \frac {\partial(\eta_1,\eta_2,.......,\eta_n)}{\partial (\zeta_1,\zeta_2 , ....., \zeta_n)} $ and $ det(J) = 1  $.

 The canonical transformation is compatible with the dynamics such that $\frac {d(\eta,\zeta)} {dt}  = J \nabla H $  is a 4nx1 matrix, and H is the Hamiltonian.The  particular choice of a non canonical transformation ${ \zeta}' =   A{ \zeta}$.\qquad

  $ { \Omega}'=   A^{T}{ \Omega A} $ =
  \begin{displaymath}  
 \left(\begin{array}{c|c}
J & O \\
\hline 
 O  & -J
\end{array} \right)  
\end{displaymath} \qquad

 inverts the sign for the n+1 to 2n variables of the p type.

The variance is  $ {V_ {ab}} = <( \bigtriangleup \zeta_a ,\bigtriangleup \zeta_b)>$. The average is with respect to the Wigner function. The symmetrised product of the deviations is taken .The $ \bigtriangleup \zeta_a $ is the deviation $(\zeta_a - <\zeta_a>)$.  The variance matrix has the corresponding transform $ V'$  = $ A^T V A  $ induced on it.

 The property $ V > 0$ and $ V +\frac{i}{2} \Omega  \geq 0 $, as a uncertainty relation,remains invariant.

 The identification of the 2n mode density matrix to the Wigner function is made as $
\ W(\zeta) = (2\pi)^{2n} (| detV | )^{-\frac{1}{2}} exp(\frac{-1}{2}\zeta^T V^{-1} \zeta)  $  following ( Ref.1) in the two mode case.

For the trivial case of diagonal Variances and Gaussian $ W(\zeta)$, the product of the exponentials is a exponential with sum of exponents ,all of gaussian form. Hence factorisability is automatic. A Hermitian variance matrix with real eigenvalues, then ensures the positivity (non negativity ) condition on the coeffecients of the density matrix.

The Schmidt decomposition of the density operators is: $ \rho  = \sum_j p_j \rho_{j1}\bigotimes \rho_{j2} $. For non negative $ p_j$, in the bipartite case, it is the neccessary and sufficient condition for separability. The particular choice of pairwise reflection on half the congugate variables, then extends the resulting condition for separability of the two mode case to the 2n mode case.
}
\section{}

{Discussion of the Conditions for Separability

Consider the first  n variables as  one system and the second half variables as the other system. It is in the latter that the mirror inversion is applied to each one of the pair of congugate variables ( say, the momentum like ones) as they occur in the Wigner function.

 The partial transpose operation is so defined. In the definition of the density matrix above the index  j runs over 1 to n for system one and n+1 to 2n for system two. These are often called system A (Alice) and B (Bob); although unlike in quantum information theory , they are better seen as a particular partition of the original system for many particle classical and quantum dynamics.

The correspondence between classical phase space and quantum phase space can now be made. Classically the separability criterion  involves  finding a canonical transform  $ A\zeta $ such that the  independent normal modes of the system are obtained. In quantum dynamics 2n oscillators with their gaussian ground states have been made independent degrees of freedom.

 Classically model Hamiltonians that admit  a reduction to integrable and separable variables,could be used, at least in subspaces of the same dimension in phase space (the 2n torii). The Kolmogorov, Arnold and Moser theorem  provides this. Quantum mechanically, the use of non gaussian functions , can arise for general hamiltonians for the 2n + 2n variables.

 However then, the absence of entanglement or criterion for separability is  not automatically true as a neccessary condition. A model dependent analysis may be needed to obtain the examples of nonentanglement or separable systems.

The question remains open about systems with $ ( 2n + 1) + ( 2n + 1) $ variables. The two mode analysis, which was generalised to the 2n mode one, is known to have problems even for 2x3 systems. For higher dimensions than three, the separability condition is not sufficient.

 For tripartite systems separability for GHZ and W states exists,these are only two inequivalent classes of states. There are results for biseparable cases with a pair x one condition ; but no general result is known.The methods developed here could be applied to the condition of biseparatibility.

In conclusion it is likely that entanglement and non separability is the generic case . Although a significant number of interesting or relevant examples of quantum dynamics with separability may also exist. The parallel situation in classical dynamics is that a significant number of integrable examples exist, as does chaos generally. Both bound and free solutions to the classical and quantum dynamics have conditions of separability.

In quantum dynamics the explicit calculation of the Wigner and related distributions while known analytically for some cases ,can give as in quantum chaos, numerically computed and graphically plotted Husimi distributions, showing fragmented  or multiply connected phase spaces. Many of the features are ``model dependent `` on a class of Hamiltonians.
}

\section{}

{Applications to Common Systems with Entanglement Possibilities

The separability criterion for quantum dynamics is a relatively new subject , although its likely occurence was noticed since the work of Schroedinger, Heisenberg and Dirac on foundations of quantum theory. For example for bound states in molecular bonds, the sum and product of the atom basis functions are used to explain the bond formation and orbitals for electron states.

 Measures of entanglement and separability have significance here. In many particle systems in which quasi particle excitations give rise to a reduced quantum mechanics ; correlation and entanglement effects are being discovered for example in spin chains. It is likely  leading to a reinterpretation of previous results and paradigms of condensed matter physics.

While the classical phase space has the symplectic group structure , the non commutative quantum phase space requires the Heisenberg group. The Fourier Wigner distributons form a representation for the Heisenberg group that leaves the commutator algebra invariant. These are symmetric on the position like $(q_i)$ and momentum like $(p_i) $ variables.

 The general element is $ W(f,g)(x,y) = \int dqdp exp(ixq + iyp + iqp)f(q,p)g(q,p) $ in which the variables q,p;x,y are n dimensional vectors on the phase space and its dual. The previous construction could have been done with any ordered set of n variables reversing sign, for example $ q_{n+1}$ to $ q_{2n}$, instead of $ p_{n+1} $ to $ p_{2n} $.

 The interpretation of position like and momentum like variables is not maintained under the group transformations ,except that the commutator $ \left[ q_i ,p_j \right]$ equals $ \delta _{ij} i\hbar $ is maintained.
}
\section{}
{Application of Separability Conditions in Exotic Systems

Quantum symplectic geometry and non commutative geometry have a correspondence. A 2n dimensional space is used with variables $ x_i $ and $x_j $, with $\left[x_i ,x_j\right] = i \Theta _{ij} $. A suitable skew symmetric two form $ \Theta $ could be used to describe the symplectic structure on the n variables $q_i $ and  $p_i$ of a 2n dimensional phase space.

 This problem could be mapped into the Weyl group representations instead of the Symplectic/Heisenberg groups. The Moyal bracket, instead of the Poisson bracket is used. The generator is a exponentiated skew form $ exp (i\frac{\partial}{\partial x_i}\Theta_{ij} \frac{\partial}{\partial x_j})$. This makes the entanglement conditions more complicated

There is also a correspondence to quantum groups , with ``Q'' deformed algebras on the phase space variables. The deformed oscillator representations are known in quantum optics. They are generated by the  hermitian linear combinations of the products of the operators  $a_i $ and $ a_i^+ $ . These are written in terms of the phase space variables ; which  induce commutators : $a_i a_j^+ - Q a_j^+ a_i =\delta_{ij}$.

 These representations are similar to the squeezed state representations , expressed in terms of Wigner like functions. Supersymmetric quantum mechanics has been expressed in terms of quantum groups ; as well as the interpolating and fractional, general  spin states. Questions of separability and entanglement in the Hilbert space arise for them too.

The extension to Bosonic and Fermionic variables requires respectively commutator and anticommutator algebras on phase space ; and this has been achieved with Grassmann algebra as well as with non commutative geometry , even in quantum mechanics. The separability conditions for Bosonic variables are likely to be similar to the oscillator state conditions, with Wigner distributions and the variance conditions for Schmidt decomposition.

 For the Fermionic system ``phase space'', the construction of Wigner like distributions and their application are also a subject of current research.Entanglement of spin and phase space degrees of freedom could be formulated ; with projection to the spin and momentum sector for helicity measurements , or to the spin and position sector for real space effects. 
}
\section{}
{Phase Space Methods and Field Theory

The possibility of infinite dimensional representations of the Heisenberg group and Wigner distributions for fields exists. There exist corresponding relations in finite dimensional systems which are generalised. The coherent state representations of the Wigner distributions on phase space are a continuum. The displaced oscillator unitary operators are used to generate these in quantum optics.

 For  the n oscillator systems,  including  generalisations , the complex (Kahler) manifold representation of the Heisenberg group algebra can be used. It would be interesting to adapt the functional integral methods  for phase spaces of field theory. The n point Green function factorisable as two point functions, and expressed as Gaussian integrals is a special case of separability.

 There are known symplectic structures on the phase space of gauge fields and potentials. Non commutative geometry effects in gauge field theory are known. The phase space of Yang Mills  gauge fields has moduli space due to the action of the gauge group. Gravity is also a gauge field theory in Ashtekar variables. It is complicated to quantise these theories and loop space holonomy representations are used to simplify the path integrals and partition function calculations over underlying spaces.

 The  Wigner distributions on phase space (modulo gauge) may provide a representation of the quantum theory of the gauge field  observables,  using classical functions on phase space. This has the advantage of'' smoothing out'' the  calculated quantities ; although singular and non positive functions can also occur.

 What are the prospects of understanding separability conditions, and interpreting the effects, for non abelian gauge field theory done with phase space methods and Wigner functions? Some indicators come from scalar and electromagnetic field theories done this way, for example in quantum optics. Abelian gauge field theory on non commutative geometry is being mapped onto non abelian gauge field theory on commutative geometry, with doubling of the variables (dimension). The symplectic , non abelian group and non commutative geometry structures of the theory are getting correlated.

Work is going on to find the correspondences among these new developments of mathematical physics . The methods of quantum and classical phase space  continue to play an important role in fundamental physics.  
}
\section{}
{Acknowledgements

The author thanks the Institute of Mathematical Sciences ,Chennai, India ; where this work was written, for hospitality, facilities and discussions. 
}
\section{}
{References

(1) R.Simon,v.84,\# 12 , 2726 -2729 , Physical Review letters(2000)

(2) P.Horodecki , Phys Lett A 232 , 333 (1997)

(3) A.Peres, PRL 77,1413,(1996)

(4) J.Jose, Classical Dynamics, Cambridge University Press (1999)

(5) S.Stenholm, Journal of Modern  Optics (2000),v47,\#2/3,311-324, special issue on Quantum Information and computation

(6) D.Bruss, Journal of mathematical physics, September( 2002),v43\#9, 4237-4251, special issue on Quantum Information Theory

(7) M.Keyl, Physics Reports, 369(2002), special issue on Fundamentals of Quantum information theory

(8) Classical and quantum systems Ed H.D.Doebner , W.Scherer , F. Schroeck, World Scientific Press( 1993),Proc of IInd International Wigner symposium 

(9) G. Folland, Harmonic analysis in phase space,Princeton University Press

(10) Kim and Noz, Quantum mechanics on phase space, (WSP)

(11) M.Henneaux, C.Teitelboim, Quantisation of Gauge systems; Princeton University press(1992).

(12) J. Madore, Noncommutative geometry and applications in physical sciences(Cambridge University Press, 1999)

(13) D. Kastler and A.Connes, J.Mathematical physics v41,\#6 June( 2000) special issue on noncommutative geometry

(14) Lectures on Quantum Mechanics and index theorem, Orlando Alvarez, Geometry and Quantum Field theory( 1995), ed .Daniel Freed and Karen Uhlenbeck

(15) J.Madore, S.Schraml ,P.Schupp, J.Wess, Gauge theory on non commutative spaces. hep-th/0001203

(16) Robert Oeckl, The quantum geometry of spin and statistics. hep-th/0008072

(17) The geometry of Quantum mechanics , Jose Isidro , hep-th/0110151

}
\end{document}